\documentclass[12pt]{article}
\usepackage{graphicx}
\usepackage{color}

\def\bbeta{\mbox{\boldmath$\beta$}}
\begin{document}
\centerline{\bf  NEW ITEM RESPONSE MODELS}
\centerline{\bf  APPLICATION TO SCHOOL BULLYING DATA}
\centerline{\bf Edilberto Cepeda-Cuervo}
\centerline{email: ecepedac@unal.edu.co}
\centerline{\bf Statistical Department}
\centerline{\bf National University of Colombia}

\section*{Summary}
School bullying victimization is a variable that cannot be measured directly. Taking into account that this variable has a lower bound, given by the absence of bullying victimization, this paper proposes IRT logistic models, where the latent parameter ranges from $0$ to $\infty$ or from $0$ to a positive real number $R$, defining the IRT parameters and proposing an empirical anchor procedure. As the academic abilities and the school bullying victimization can be explained due to associated factors such as habits, sex, socioeconomic level and education level of parents, IRT regression models are proposed to make joint inferences about individual and school characteristic effects. Results from the application of the proposed models to the Bogot\'a school bullying dataset are presented. The need for testing based in statistical models increases in different fields.

{\it Key words: logistic models, ability modeling, item response theory, bullying data}

\section{Introduction}

In the social sciences, the research interest usually includes characteristics (such as cognitive development, quality of life or level of victimization), which cannot be measured directly (Wilson, 1989; Goodboy and Martin, 2015). The information for the study of these variables is collected using instruments (questionnaires or tests) to observe people's behavior/performance, in which questions related to the variable of interest are asked, for which there is no defined and interpretable scale in advance  (Bartholomew, 1980, 1984).

To analyze and to model these unobservable variables from observed variables, latent structure models were proposed (Lord, 1950,
Lazarsfeld, 1955; Birnbaum, 1968), also known as item-response models (Mu\~{n}iz,  1997). These models seek to clearly represent the relationship between the various reactions, items or behaviors observed with the unobservable variable of interest, also called the latent variable. Specifically, the item-response theory (IRT) proposes a relationship between an unobserved variable (usually called ability) and the probability of a person correctly responding to any item of a test (Lord, 1980). The best known IRT models are the one-parameter, two-parameter and three-parameter logistic models. All of them assume that the underlying unobserved variable (latent variable) ranges from $-\infty$ to $\infty$, although in practical analysis it is assumed to take values  from $-3$ to $3$, with ability scaled to have mean $0$ and standard deviation 1 (Harris, 1989). Other researchers assumed that the latent variable ranges from $-4$ to $4$ (Ludlow and Haley, 1995). These IRT  logistic models have  an item difficulty parameter, corresponding to the inflection of the item's characteristic curve, the curve of the probability of correctly answering the item as function of ability. This parameter ranges theoretically from $-\infty$ to $\infty$ (or practically from $-3$ to $3$ or $-4$ to $4$). In these models, the parameter estimates are usually obtained under the assumption that subjects' abilities are random samples from an $N (0, 1)$ ability distribution (Bock and Lieberman, 1970). The parameter estimates can also be obtained, freeing the method from arbitrary assumptions about the distribution of ability in the population sampled (Bock and Aitken 1981), estimated as a discrete distribution at a finite number of points (i.e., a histogram).

The need for testing based in statistical models increases in different fields.
Item-based statistical models are used to measure the quality of education, development of communicative skills and mathematical ability, among others. In this paper, we use these models to measure the intensity of school bullying victimization. However, since there is a  lower bound of victimization given by absence of bullying and an upper bond assumed by the test, an IRT with latent parameter ranging from $\infty$ to $\infty$ is inconsistent with the framework of this application. As a consequence, a new model specification should be considered to take into account the framework of the test and to determine the true distribution of the latent variable (e.g. ability or bullying victimization). For this reason, in this paper we propose new IRT models where the range for the individual ability is from $0$ to $\infty$, and from $0$ to $R$, where $R$ is a positive real number. In the first case, the model is defined assuming logarithmic transformation for the abilities and in the second, assuming a logit, probit or log-log transformation, after rescaling the abilities from the interval $(0,R)$ to the interval $(0, 1)$. This new range of the latent variable can be more appropriate to study levels of bullying victimization. Thus a special modeling approach is considered under a Bayesian perspective, where uniform prior distributions are assumed for the difficulty and subject latent parameters. Additionally, this paper presents the anchoring procedure of the scale introduced by Beaton and Allen (1992) and a proposal for empirically obtaining the performance levels, which allows the interpretation of the skill scale.

The  proposed  IRT  models  were  fitted  to  Bogot\'a  school  bullying  data, finding that all capture the skewness of the latent parameters (bullying victimization). After the estimation of the parameters involved in the models and the anchoring processes of the items (Beaton and Johnson, 1992; Andrade et al., 2000), the values of the scale that allow assigning a substantive interpretation with respect to the construct (bullying victimization) is appropriately selected. Also, since bullying victimization can be explained due to associated factors such as habits, sex, socioeconomic level and education level of parents, IRT regression models are proposed and applied to analyze the Bogot\'a bullying data.

The organization of the paper is as follows. After a brief introduction, in Section \ref{3IRT}, the paper presents some topics of three-parameter logistic models where the range for the individual latent parameter (ability) is from $0$ to $\infty$, and from $0$ to $R$, where $R$ is a positive real number. In Section 3, the ability scale is presented. Modeling of abilities is presented in Section 4. Results of the Bogot\'a bullying data analysis are presented in Section 5. Section 6 concludes.

\section{Three parameter logistic models \label{3IRT}}

The objective of this section is to propose new item-response models to study the school bullying victimization phenomenon. The proposed models present a good alternative to analyze sets of dichotomic items taking into account theoretical frameworks of educational and psychological tests, where the latent and difficulty parameters range in the positive real number set or in a positive bounded interval of the real numbers. These models provide estimates of the parameters in a scale with easy interpretation and understanding by the professionals who construct, apply and interpret questionnaires.

\subsection{IRT model with abilities ranging in the real number set.\label{3nomMod}}
In this section, the traditional 3-IRT model is presented.
Suppose that $u_{ij}$, $i=1,...,I$, and $j=1,...,n$, are $n\times I$ binary random
variables, where  $i$ indicates an item and $j$ indicates a subject. $u_{ij}=1 $ if subject $j$
solves item $i$ correctly, and $u_{ij}=0 $ if not. In this model, the probability that a  subject $j$
with ability  $\theta_j$ solves item $i$ correctly, with difficulty parameter $b_i$, is:

\begin{eqnarray}
P(u_{ij}=1|\theta_j, \xi_i)=c_i+(1-c_i)\frac{1}{1+e^{-Da_i(\theta_j-b_i)}},
\mbox{   }   i=1,...,I,  j=1,...,n,\label{m3irt}
\end{eqnarray}
where $\xi_i=(a_i, b_i, c_i)$ are the parameters of  item $i$, $0\leq c_i<1$ and $a_i>0$ (Birnbaun, 1968). $\theta_j$ and  $b_i$ assume values between $-\infty$ and $\infty$. The latent continuum $\theta$ is called ability, and $\theta_j$ is the ability of the j-th subject. $D$ is a  constant which can be arbitrarily set. It is customary  to set $D=1.7$.

The following considerations provide a basis for interpreting the parameters involved in the IRT model:

\begin{enumerate}
\item As $c_i=\displaystyle\lim_{\theta_j\to -\infty}P(u_i=1|\theta_j, \xi_i)$, $c_i$ can be
defined as the probability of random correct answers.

\item If $\theta_j=b_i$, from
equation (\ref{m3irt}),

\begin{eqnarray}
P(u_{ij}=1|\theta_j, \xi_i)=\frac{1+c_i}{2}\label{eq2}.
\end{eqnarray}
Thus, $b_i$ can be interpreted as the ability necessary for  a subject to solve  item
$i$ with probability equal to $\frac{1+c_i}{2}$. As a consequence  of equation \ref{eq2}, for the same probability
of random answer $c_i$, large  values of $b_i$ correspond to
items with large difficulty, given that items with larger  difficulty values need larger
ability to have a correct answer probability  equal to $\frac{1+c_i}{2}$.

\item Assuming that $P(u_{ij}=1|\theta_j, \xi_i)$  is a function $f(.)$ of $\theta_j$, the first order derivative of
$f$ evaluated at $b_i$ is given by:
\begin{eqnarray}
f'(b_i)=\frac{1}{4}(1-c_i)Da_i.\nonumber
\end{eqnarray}
Thus, given $b_i$ and $c_i$, $a_i$  can be assumed as a discrimination measure. If $c_1=c_2=c$ and $a_1<a_2$, an
item with  parameters $c$ and $a_2$  discriminates more than an item with  parameters $c$ and $a_1$.
\end{enumerate}

In equation (\ref{m3irt}), if $c_i=0$, there is no random possibility of correct
answer and the resulting model is called a unidimensional logistic model for two
parameters or Birnbaum's model (Van der Linden and Hambleton, 1997).
Setting $a_i = a$ and $c = 0$ in equation (1), one obtains the unidimensional one
parameter logistic model, also called the Rasch model (Rasch, 1961). The Rasch model
describes how the probability of correct answers depends on the subject's overall
ability and the level of difficulty of the questions.

One of the most important theoretical merits of the Rasch model is the so-called ``specific
objectivity" (Fisher, 1995). This means that the item parameters do not depend on the characteristics of the persons answering the test, and that the personal parameters do
not depend on the items, specifically chosen from a set of items. Consequently, the
three parameters are independent of the average ability of the sample  and its respective dispersion.

\begin{figure}[h]
\centering
\includegraphics[scale=0.4]{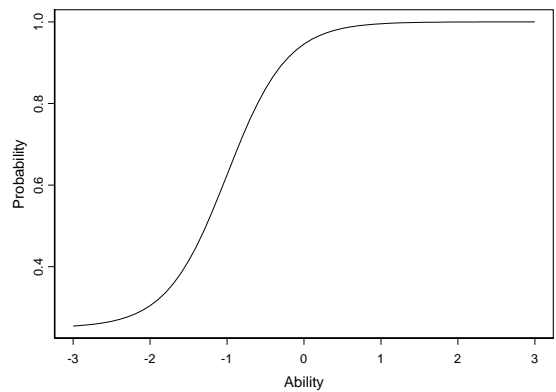}
\caption{Item Characteristic Curve   a=1.5, b=1, c=0.25, D=1.7. }\label{nf3ir}
\end{figure}

Given an item and known parameters, the curve which describes the relation of subjects' ability
parameter and agreement probability  is called the Item Characteristic Curve (Hambleton et al., 1991).   Figure \ref{nf3ir} shows this curve for an item with
$D=1.7$ and parameters $a=1.5$, $c=0.25$ and $b=1$. The curve shows that the probability of a correct answer $P(u_{ij}=1|\theta_j)$ increases when $\theta$ increases. This means that, for example, in mathematical performance, the probability of correct response increases with mathematical ability.

\subsection{ New IRT response models}
\subsubsection{ IRT with abilities ranging from $0$ to $\infty$.} \label{IRTLog}

Given that the bullying victimization, communicative competence or the  school math performance of the student is not zero and can always be improved, in this paper we assume  that the ability parameter
takes values in the positive real numbers. In this model, denoted 3-IRT,  the probability that a
subject $j$ with ability parameter $\theta_j$ solves an item
$i$ with difficulty parameter $b_i$ is given by:
\begin{eqnarray}
P(u_{ij}=1|\theta_j, \xi_i)=c_i+(1-c_i)\frac{1}{1+e^{-Da_i(\log(\theta_j)-\log(b_i))}},\label{m3irtp}
\end{eqnarray}
$ i=1,...,I,\;  j=1,...,n$,  where $\xi_i=(a_i, b_i, c_i)$ are the parameters of item $i$, $0\leq c_i<1$ and $a_i>0$. Here  $\theta_j$ and $b_i$ assume values between $0$ and $\infty$. The latent continuum $\theta$ is called the ability, and $\theta_j$ is the ability of the j-th subject.
As $c_i=\displaystyle\lim_{\theta_j\to 0^{+}}P(u_i=1|\theta_j, \xi_i)$, $c_i$ can be
defined as the probability of random correct answers. Also, if $\theta_j=b_i$, from
equation (\ref{m3irtp}) it follows that:

\begin{eqnarray}
P(u_{ij}=1|\theta_j, \xi_i)=\frac{1+c_i}{2}.
\nonumber
\end{eqnarray}
Thus, $b_i$ can be interpreted as the subject's ability necessary to solves an item
$i$ with probability equal to $\frac{1+c_i}{2}$. Large  values of $b_i$ correspond
to large item difficulty.

Assuming $P(u_{ij}=1|\theta_j, \xi_i)$ is a function $f(.)$ of $\theta_j$, the first-order derivative of $f$ relative to $\theta_j$, evaluated  on $b_i$, is given by:
\begin{eqnarray}
f'(b_i)=\frac{1}{4}(1-c_i)Da_i/b_i.\nonumber
\end{eqnarray}
Thus, given $c_i$ and $b_i$, $a_i$  can be treated as a discrimination measure. If $c_1=c_2$, $b_1=b_2$ and $a_1<a_2$,  where the subscript is the item number, then item $2$ discriminates more than item $1$ between subjects with different ability levels. Figure 2 shows the  characteristic curve for items with
with $D=1.7$ and item parameters  $(a,b,c)=(3, 1, 0.25)$, on the left, and  $(a, b, c)=(2.5, 1.5, 0.25)$, on the right.

\begin{figure}[h]
\centering
\includegraphics[scale=0.34]{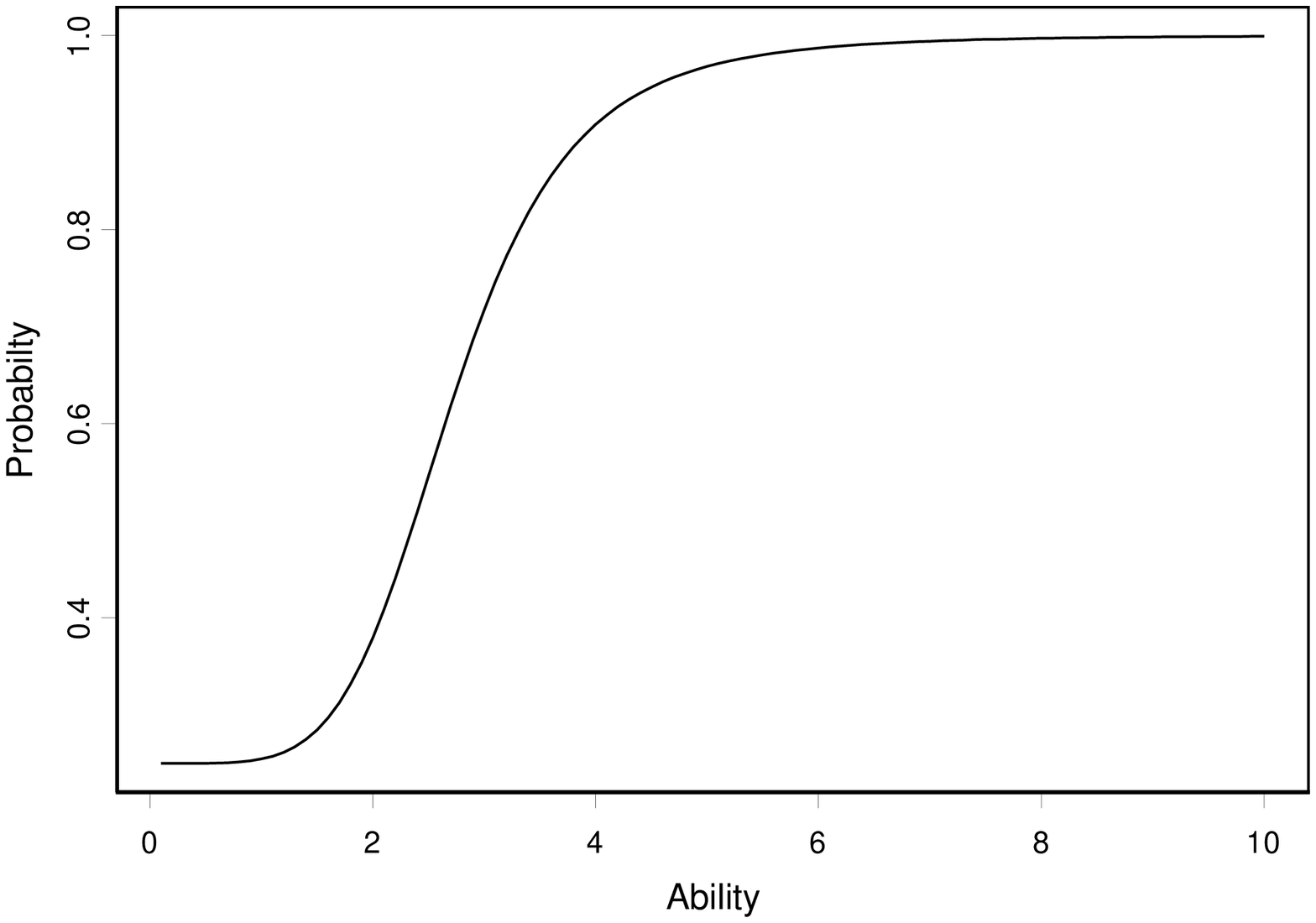}
\includegraphics[scale=0.34]{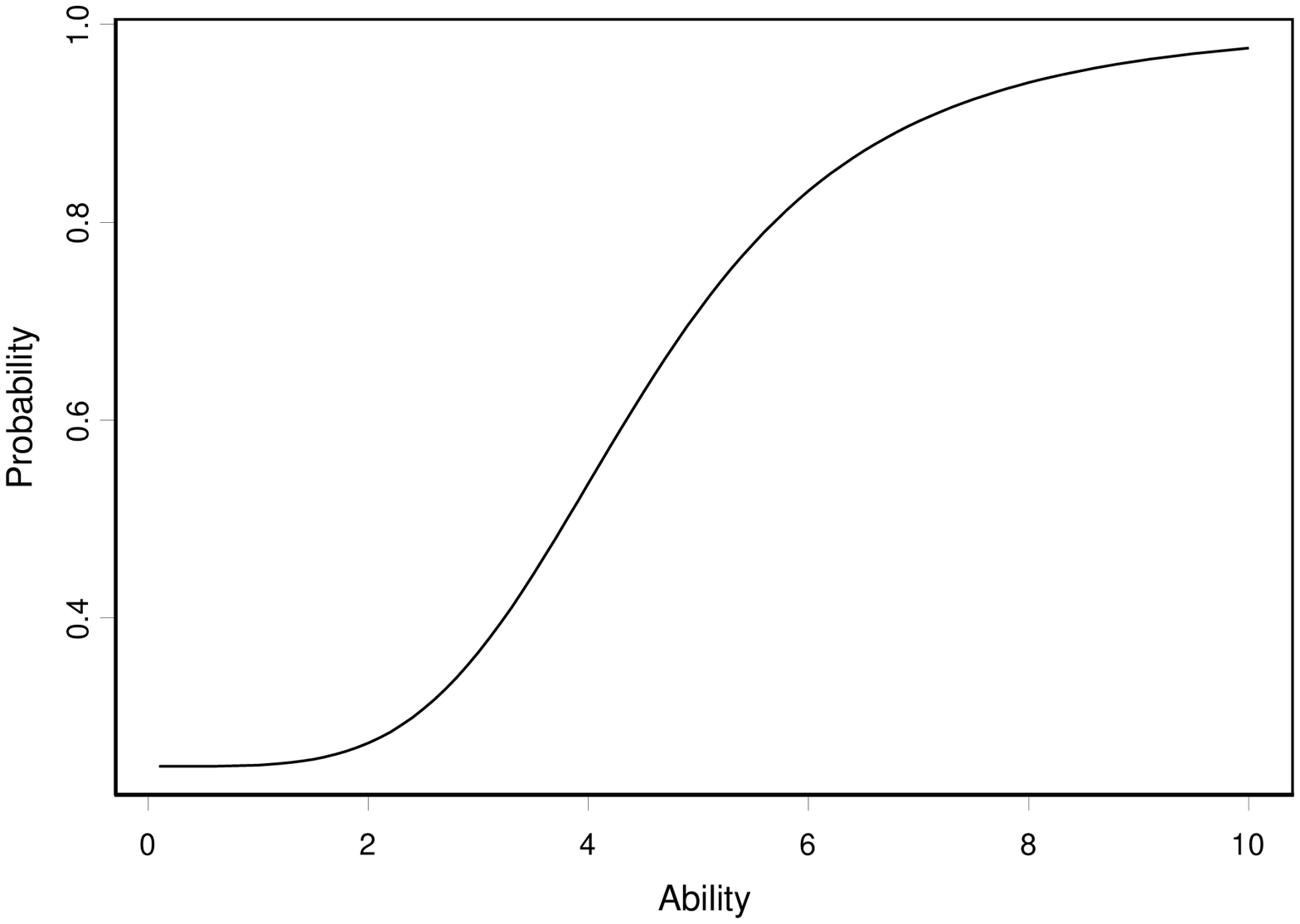}
\caption{ Item Characteristic Curve under a=3 , b=1 and c=0.25, on the left, and under  a=2.5, b=1.5 and c=0.25,  on the right.}
\end{figure}

In equation (\ref{m3irtp}), if $c_i=0$, there is no random probability of correct
answers and the resulting model is called the unidimensional logistic model for two
parameters or the L-Birnbaum's model. If $a_i=a$ and $c=0$, one obtains the unidimensional logistic model of
one parameter, called the L-Rasch model. The L-Rasch model describes how the probability of correct responses depends on the subject's overall
ability and the level of difficulty of the questions.

In the usual model definition, equation 1, the discrimination parameter does not depend on the difficulty parameter. However, in the  proposed model the discrimination depends on the difficulty parameter.

\subsubsection{IRT model with abilities ranging in a bounded interval} \label{IRTUnif}

The  three-parameter uniform model, denoted 3-CIRT,  is defined as a three-parameter normal model, assuming that the
probability that a  subject $j$ with ability  $\theta_j$ solves  item $i$, with difficulty parameter $b_i$, is given by the equation:

\begin{eqnarray}
P(u_{ij}=1|\theta_j, \xi_i)=c_i+(1-c_i)\frac{1}{1+e^{-Da_i(\mbox{logit}(\theta_j/R)-\mbox{logit}(b_i/R))}},\label{m3uirtp}
\end{eqnarray}
$i=1,...,I,\;  j=1,...,n$, where $\xi_i=(a_i, b_i, c_i)$ are the parameters of item $i$, $0\leq c_i<1$ and $a_i>0$. Here  $\theta_j$  and $b_i$  assume values in the open interval $(0, R)$. The latent continuum $\theta$ is  the ability, and $\theta_j$ is the ability of the j-th subject.
As $c_i=\displaystyle\lim_{\theta_j\to 0^{+}}P(u_i=1|\theta_j, \xi_i)$, $c_i$ can be
defined as the probability of random correct answers. Also, if $\theta_j=b_i$, from
equation (\ref{m3uirtp}) it follows that:

\begin{eqnarray}
P(u_{ij}=1|\theta_j, \xi_i)=\frac{1+c_i}{2}.
\nonumber
\end{eqnarray}
Thus, $b_i$ can be interpreted as the subject's ability necessary to solves an item
$i$ with probability equal to $\frac{1+c_i}{2}$. Large  values of $b_i$ correspond
to large item difficulty.

Assuming $P(u_{ij}=1|\theta_j, \xi_i)$ is a function $f(.)$ of $\theta$, the first-order derivative of $f$ relative to $\theta$, evaluated  on $b_i$, is given by:
\begin{eqnarray}
f'(b_i)=\frac{1}{4}(1-c_i)Da_i\frac{R}{(R-b_j)b_j}.\nonumber
\end{eqnarray}
Thus, given $c_i$ and $b_i$, $a_i$  can be treated as a discrimination measure. If $c_1=c_2$, $b_1=b_2$ and $a_1<a_2$,  where the subscript is the item number, then item $2$ discriminates more than item $1$ between subjects with different ability levels: the discrimination depends on the difficulty parameter. Figure 2 shows the  characteristic curve for the uniform item response model.

In equation (\ref{m3uirtp}), if $c_i=0$, there is no random possibility of correct
answers and the resulting model is called the unidimensional logistic model for two
parameters or the C-Birnbaum model. If $a_i=a$ and $c=0$, one obtains the unidimensional logistic model of
one parameter, called the C-Rasch model. The C-Rasch model describes how the probability of correct responses depends on the subject's overall
ability and the level of difficulty of the questions.

\begin{figure}[h]
\centering
\includegraphics[scale=0.34]{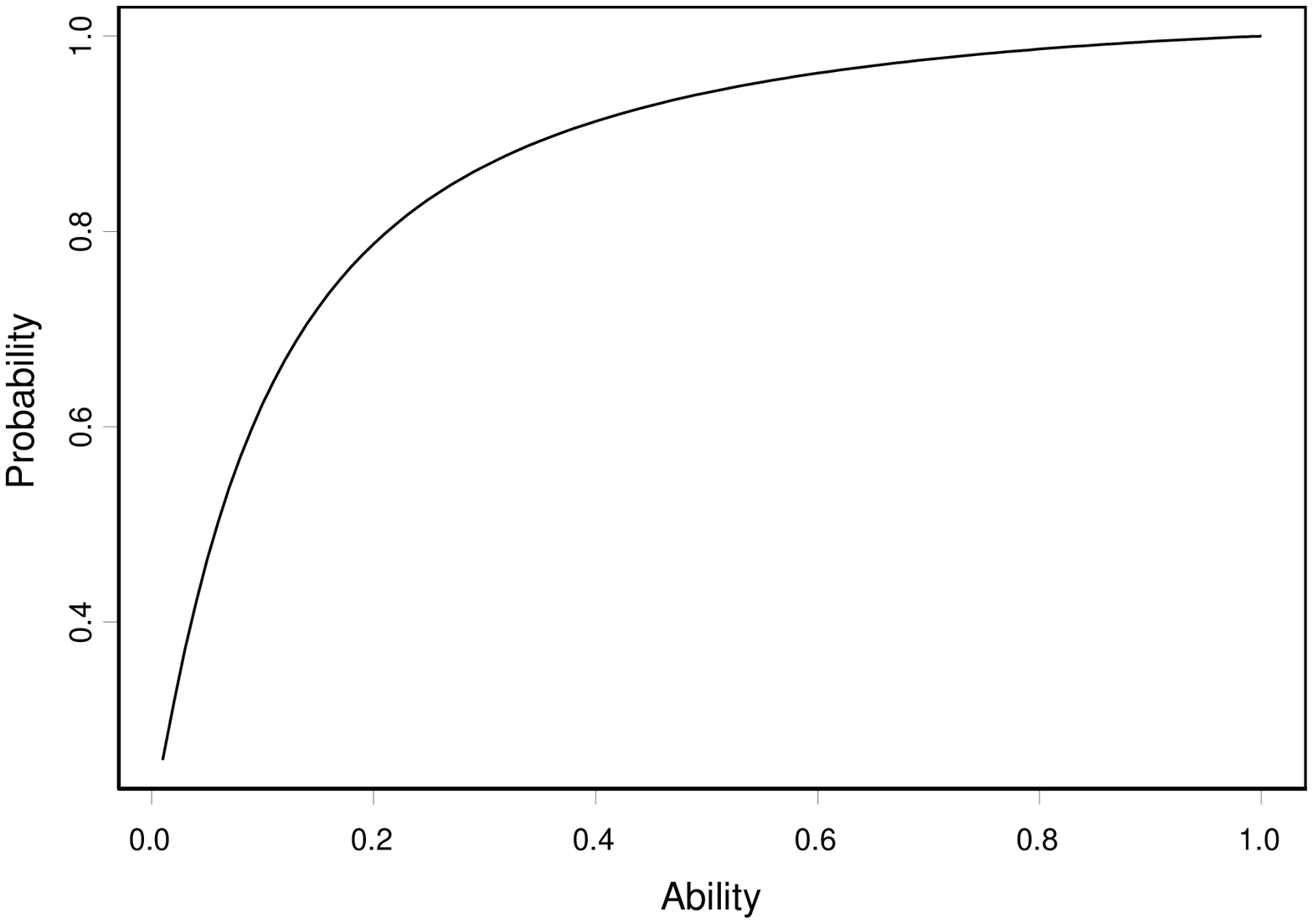}
\includegraphics[scale=0.34]{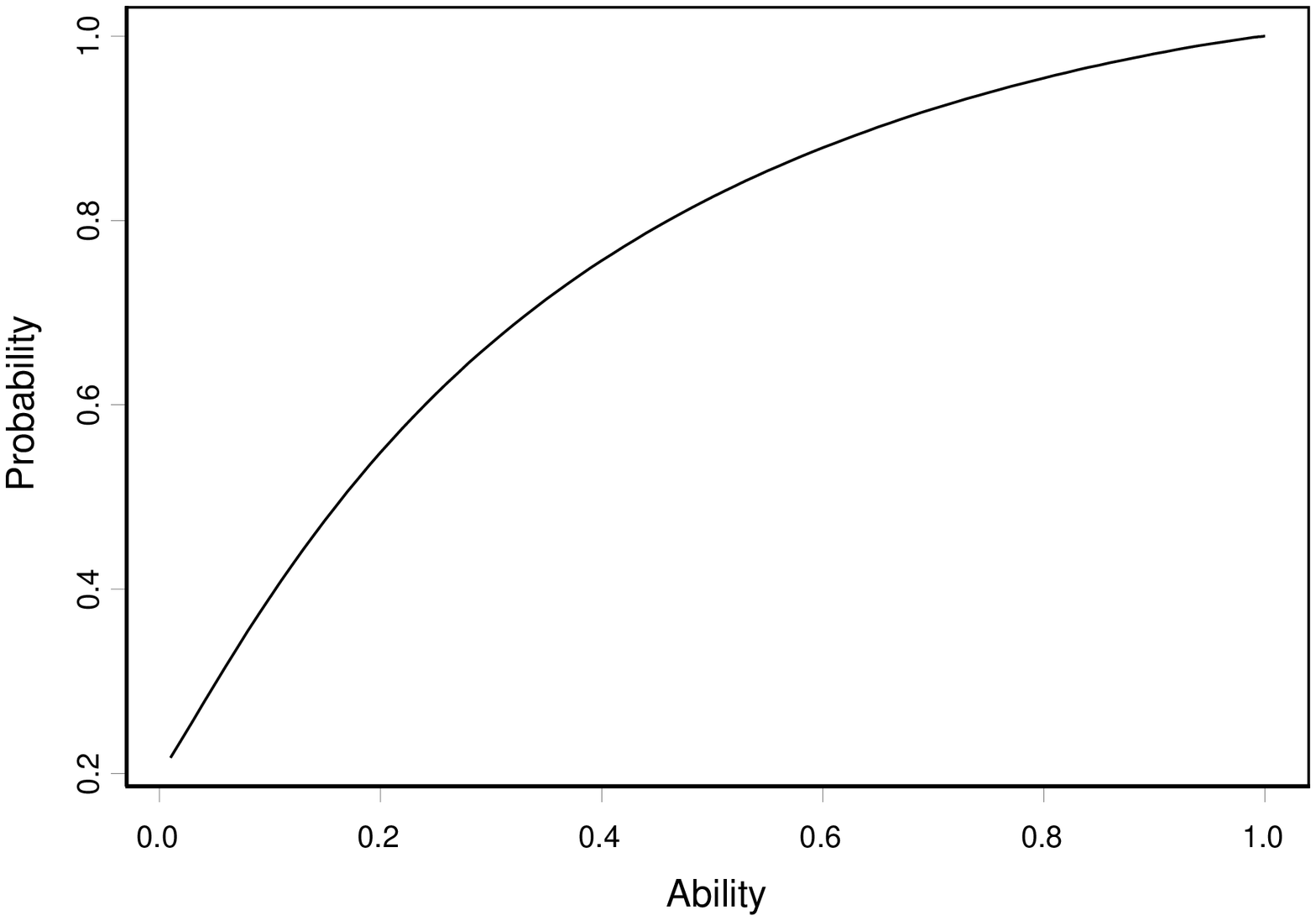}
\caption{ Item Characteristic Curve under $a=1.5$ , $b=-0.5$, $c=0.25$ and $D=1.7$, on the left, and under  $a=1.5$, $b=-1$, $c=0.25$ and $D=1.7$,  on the right.}
\end{figure}

In the 3C-IRT model, equation \ref{m3uirtp}, there are other possible choices of the transformation functions, different from  the logit specification, $\mbox{logit}(\theta/R)=\log(\theta/(R-\theta)$). Examples are the probit function
$g(\theta)=\Phi^{-1}(\theta/R)$, where $\Phi(.)$ is the cumulative distribution function of a standard
normal random variable; the complementary log-log function $g(\theta)=\log(-\log((R-\theta)/R)$, and
the log-log function $g(\theta)=-\log(\log(R)-\log(\theta))$, among others.

\section{Ability scales\label{ability}}

Since the ability scale does not ``itself" have  practical interpretation (e.g., in an educational
context), it is necessary to define a practical ability scale,  characterized by sets of items
that give an interpretation  (e.g., pedagogical) in the theoretical framework of the test. The
ability scale is defined by a set of $values$ $\theta_0, <\theta_1<...<\theta_P$
of $\theta$, called ``anchor levels", selected by the analyst (Andrade et al., 2000).
In our applications, the interpretation of the scale, determined by bullying victimization levels, is possible across the
psychological interpretation of the items that are associated with each level.

The  pertinent anchor levels, $\theta_p$, $p=1,2,\dots,P$, depend on the conditional probabilities
of a correct answer $P(u=1|\theta=\theta_p)$  and  $P(u=1|\theta=\theta_{p-1})$ (Beaton and Allen, 1999;  Andrade, 2000).
Specifically, an item $u$ is an anchor item of a level $\theta_p$, if and only if,
\begin{enumerate}
\item $P(u=1|\theta=\theta_p)\geq 0.65$,\label{cond1}
\item $P(u=1|\theta=\theta_{p-1})<0.5$ \mbox{  and }\label{cond2}
\item $P(u=1|\theta=\theta_p)-P(u=1|\theta=\theta_{p-1})\geq 0.30$\label{cond3},
\end{enumerate}
where values $\theta_p$ and $\theta_{p-1}$ are related to conditions \ref{cond1} and  \ref{cond2}  and the  third condition is related to the  difference between the two previous probabilities.

In this paper, an empirical procedure is assumed for the selection of skill levels for a set of items. In this method, presented in Cervantes et al. (2008), for an  item it is possible to search for a pair of points $(\theta_{\ell}, \theta_h)$ in the skill scale that meet the necessary criteria for the anchoring, such that any ordered pair
of points $(\theta_1, \theta_2)$ in which $\theta_1 \leq \theta_{\ell}$ and $\theta_2 \geq \theta_h$ would allow anchoring the item at level $\theta_2$ related to the level $\theta_1$, and where $\theta_1 > \theta_{\ell}$ and $\theta_2 <\theta_h$
 would not allow anchoring the item. The pair $(\theta_{\ell}, \theta_h)$ provides the
shortest interval in the skill scale for which it is possible to anchor the item.
For  one and two-parameter models, the pair is given by the values
of $(\theta_{\ell}, \theta_h)$  such that $P(u=1|\theta_{\ell})=0.35$ and $P(u=1|\theta_h)=0.65$, respectively.  In the three-parameter mode this interval is given by the values of $\theta$ such that:
\begin{enumerate}
\item $P(u=1|\theta_{\ell})=\frac{c+1}{2}-0.15$ and $P(u=1|\theta_h)=\frac{c+1}{2}+0.15$ if $c<0.3$
\item $P(u=1|\theta_{\ell})=0.35$ and $P(u=1|\theta_h)=0.65$ if $0.3\leq c<0.35$
\item $P(u=1|\theta_{\ell})=0.5-\epsilon $ and $P(u=1|\theta_h)=0.8-\epsilon$, if $c\ge 0.35$ and $\epsilon>0.35$.
\end{enumerate}
It  $c\geq 0.5$, no pair $(\theta_{\ell}, \theta_h)$ meets all three criteria.  Once the intervals for each items have been determined, the procedure to find $P$ and  $\theta_p$, $p=1,2,\dots,P$  such that it is  possible to include the items in $P$
levels is:
\begin{enumerate}
\item Sort the items in ascending order, according to the $\theta_{\ell}$'s.
\item Take at the lowest level $\theta_0$, the smallest  $\theta_{\ell}$'s.
\item If $p=1$, take as candidate $\theta^*$ for  $\theta_1$, the $\theta_h$ corresponding to pair $(\theta_0, \theta_h)$.
If $p=2,...,P$, take as candidate $\theta^*$ for  $\theta_p$, the $\theta_h$ corresponding to pair $(\theta_{p}, \theta_h)$.

\item Compare the candidate with the interval found for the next item:
\begin{enumerate}
\item If $\theta_{\ell}< \theta^*<\theta_h$, let $\theta^*=\theta_h$ and repeat the
comparison with the interval of the next item.
\item If $\theta^*>\theta_h$, keep the candidate and make the comparison with the following interval.
\item If $\theta^*<\theta_{\ell}$, then $\theta_p$ determine the $p$-th level of performance.
\end{enumerate}
5. Repeat steps 3 and 4 to determine the levels until having compared all the items.
\end{enumerate}

This procedure is proposed as a tool to describe the skill scale, finding points that define performance levels (Beaton and Allen, 1992). A representation of the intervals of the items, ordered according to their lower limit, allows identifying levels of achievement, determined by "jumps" between the anchor intervals. If the proposed procedure does not allow reaching skill levels, by eliminating some items of the test it is possible to define levels of performance. The interpretation of the items anchored in levels seeks to give a generalization of the abilities of the examinee population.

The procedure proposed for empirically obtaining performance levels (here bullying victimization level) allows us to obtain a closer approximation of the attribute present in the measurement instrument from the reagents that effectively compose it. This procedure allows achieving the main purposes, which is the anchorage of the item scale: 1) locating for a performance level a set of items that belongs to each of these levels. 2) facilitating the formulation of a theory about the relationship between performance levels of the attribute evaluated in the individuals and the items that compose the test, and 3) facilitating communication of the results of the evaluation to people with no expertise in the area.

\section{Modeling of abilities \label{Regress}}

Response data are often obtained in combination with other input variables. For example, response data are obtained from respondents together with school information, and the objective is to make joint inferences about individual and school effects for an outcome variable. In a Bayesian framework, different sources of information can be handled efficiently, accounting for their level of uncertainty. We will show that the flexibility of a Bayesian modeling approach together with the powerful computational methods provides an attractive set of tools for analyzing response data.

A subject's ability can be explained by associated  variables, such as: habits, practice time,  socioeconomic level and parents' educational level, using the model $\theta_j=\mathbf{x}_j^{t}\bbeta+ \epsilon$,  where $\mathbf{x}_j^t=(x_{1j},....,x_{rj})$ is an explanatory variables vector,
$\mathbf{\bbeta}=(\beta_1,...,\beta_r)$ is a regression parameters vector and $\epsilon_j\sim N(0,\sigma^2)$. Thus, model (\ref{m3irt})
can be written as:

\begin{eqnarray}
p(u_{ij}=1|\mathbf{\bbeta}, \xi_i,\epsilon, \mathbf{x}_j^t )=c_i+(1-c_i)\frac{1}{1+e^{-Da_i(\mathbf{x}_j^t\bbeta+ \epsilon-b_i)}}\label{mxr}
\end{eqnarray}
for  $i=1,...,I$ and  $j=1,...,n$. In the log-normal model   (\ref{m3irtp}) and in the 3-CIRT model (\ref{m3uirtp}), by setting $\log(\theta_j)=\mathbf{x}_j^{t}\bbeta+ \epsilon$ and $\mbox{logit}(\theta_j)=\mathbf{x}_j^{t}\bbeta+ \epsilon$ respectively,  regression  models like (\ref{mxr}) are also obtained.

In order to interpret the regression parameters, the effect of a unit change in the $k$-th covariate, from $x_{kj}$ is to $x_{kj}+1$,
is determined in model (\ref{mxr}). Assuming that $\tilde\mathbf{x}_j^t=(x_{1j},\dots,x_{kj}+1,\dots,x_{rj})$, this model can be written as:
\begin{eqnarray}
p(u_{ij}=1|\mathbf{\bbeta}, \xi_i,\epsilon, \tilde\mathbf{x}_j^t )=c_i+(1-c_i)\frac{1}{1+e^{-Da_i\beta_k}e^{-Da_i(\mathbf{x}_j^t\bbeta+ \epsilon-b_i)}}\label{mxrb}.
\end{eqnarray}
Thus, if  $\beta_k>0$, larger values of $x_{kj}$ are associated with a higher probability of the correct responses, given that $e^{-Da_i\beta_k}<1$ and  $p(u_{ij}=1|\mathbf{\bbeta}, \xi_i,\epsilon, \tilde\mathbf{x}_j )> p(u_{ij}=1|\mathbf{\bbeta}, \xi_i,\epsilon, \mathbf{x}_j)$. If  $\beta_k<0$, larger values of $x_{kj}$ are associated with a smaller probability, given that $e^{-Da_i\beta_k}>1$ and  $p(u_{ij}=1|\mathbf{\bbeta}, \xi_i,\epsilon, \tilde\mathbf{x}_j)< p(u_{ij}=1|\mathbf{\bbeta}, \xi_i,\epsilon, \mathbf{x}_j)$. In this parameter interpretation, the other covariates are held fixed.

The parameter estimates obtained from the application of these models are helpful to determine, for example, the characteristics of these systems in order to determine strategies to improve the educational quality. To study school bullying, we use these models to determine if the bullying victimization is more present in the early or late grades of secondary school, or if there are differences in the bullying victimization between boys and girls. These results allow determining and focusing prevention policies.

\section{School Bullying Application \label{bully}}

\subsection{School bullying definition}

As defined in Cepeda-Cuervo et al., (2008) and Cepeda-Cuervo and Caicedo (2012), school bullying is a type of violence manifested by physical, psychological or social assaults by classmates or teachers, repeatedly suffered by children in the school environment. To distinguish bullying from other violent actions in this environment, such as a fight between students, two characteristics identify it: The first is the intrinsic existence of a relationship of power (dominance-submission) of the aggressor over the victim. The second is that the assault situations occur repeatedly. Olweus (1994) defined bullying or victimization in the following general way: ``A person is being bullied or victimized when he or she is exposed, repeatedly and over time, to negative actions on the part of one or more other persons."

\subsection{School bullying in Bogot\'a}
Cepeda-Cuervo et al., (2008) proposed a theoretical framework and a questionnaire with 22 questions, based on the Cisneros Autotest (O\~{n}ate and Pi\~{n}uel, 2005), to inquire about the situations of harassment of students at school. Each of the items included in this questionnaire corresponds to a "harassment strategy". Each item consists of a statement and three response options (often, sometimes, never), from which the student chooses one according to the frequency with which, in each case, the harassment situation described in each of the statements happens.

The authors conducted a survey, consisting of 22 items related to
situations and behaviors that characterize school bullying, in  709
public schools on Ciudad Bol\'ivar district of Bogot\'a city. The
survey is included in the appendix.

\subsection{The sample}
The population of interest is made up of primary and middle school students of 709 public schools in Bogot\'a, Colombia, which serve more than 28,000 students. The data obtained from a sample of 80 classes, 19 of sixth grade, 12 of seventh, 12 of eighth, 10 of ninth, 10 of tenth and 17 of eleventh grade. Thus, the sample consists of 3,226 students aged between  10 and 20 years, drawn mainly from low socioeconomic households.

\subsection{The models \label{themodels}}

Each item was dichotomized, assigning $0$ if the student "sometimes or never" is a victim of the situation described by the item and $1$ if the student is a "frequent victim" of that situation. Thus, given that the probability of a random answer is $0$, the probability of victimization is assumed to follow one of the logistic functions given by:

\begin{enumerate}
\item  {\bf Model 1a}\label{norModel}
\begin{eqnarray}
& & 	\mbox{logit}(P(u_{ij}=1|\beta, \theta_j, \xi_i)) = \beta\theta_j - b_i, \label{ma1} \\
& & 	\theta_j  \sim  N(0, 1) \;\;  j = 1,...,3226; i = 1,...,22\nonumber
\end{eqnarray}
where $\theta_j$ is a measure of student victimization and $b_i$ is a measure of the ``item victimization impact".  $\beta$ can be interpreted as a dispersion parameter  (Spiegelhalter and Thomas, 2003).

\item {\bf Model 2a}\label{lognorModel}
\begin{eqnarray}
& & 	\mbox{logit}(P(u_{ij}=1|\beta, \theta_j, \xi_i)) = \beta\log(\theta_j) - \log(b_i),\label{ma2}\\
& & 	\theta_j  \sim  LN(1.64, 1) \;\;  j = 1,...,3226; i = 1,...,22\nonumber
\end{eqnarray}
where $\theta_j$ and $\beta$ can be interpreted as in  Model \ref{norModel}a, and $b_i$ is a measure of ``item victimization impact".

\item {\bf Model 3a}\label{UniforModel}
\begin{eqnarray}
& & 	\mbox{logit}(P(u_{ij}=1|\beta, \theta_j, \xi_i)) = \beta \mbox{logit}(\theta_j/5) - \mbox{logit}(b_i/5),\label{ma2}\\
& & 	\theta_j  \sim  U(0, 5) \;\;  j = 1,...,3226; i = 1,...,22\nonumber
\end{eqnarray}
where $\theta_j$ and $\beta$ can be interpreted as in  Model \ref{norModel}a, and $b_i$ is a measure of ``item victimization impact".
\end{enumerate}

\subsection{Bayesian parameter estimates \label{IRTParamEst}}

To apply Bayesian methods to obtain the posterior parameter estimates, the following prior distribution specification is assumed for all unknown parameters. For all models, a half-normal prior distribution  $HN(0,0.0001)$ is assumed for $\beta$. This represents vague prior information but constrains $\beta$ to be positive. For the difficulty  parameters $b_i's$, a normal prior  distribution  $N(0,0.0001)$ is assumed in model 1a; a lognormal distribution  $LN(1.64, 1)$ is assumed  in Model 2a, and  a uniform  prior distribution  $U(0,5)$ in Model 3a,  imposing in all cases a zero-sum constraint of $b_i's$.

The parameter estimates were obtained from 10,000  Gibbs samples  drawn, with every 5th sample recorded, after a  burn-in period of 2,000
samples.  WinBugs (Spiegelhalter et al. 2003) was used in these simulation procedures. Convergence of the Gibbs sampler algorithm was monitored using standard existing methods such as the trace plots of the simulated samples and parallel chains starting from different initial values, to provide indication of stationarity. In all applications, the posterior samples showed the same behavior for all chains, providing strong indication of convergence.

Thus, the posterior parameter estimate of $\beta$ is $\hat\beta=1.026(0.026)$, $\hat\beta=1.014 (0.026)$ and $\hat\beta=0.584(0.029)$, for Model \ref{norModel}a, Model \ref{lognorModel}a and Model \ref{UniforModel}a, respectively.

%

\begin{figure}[h!]
\centering
\includegraphics[scale=0.34]{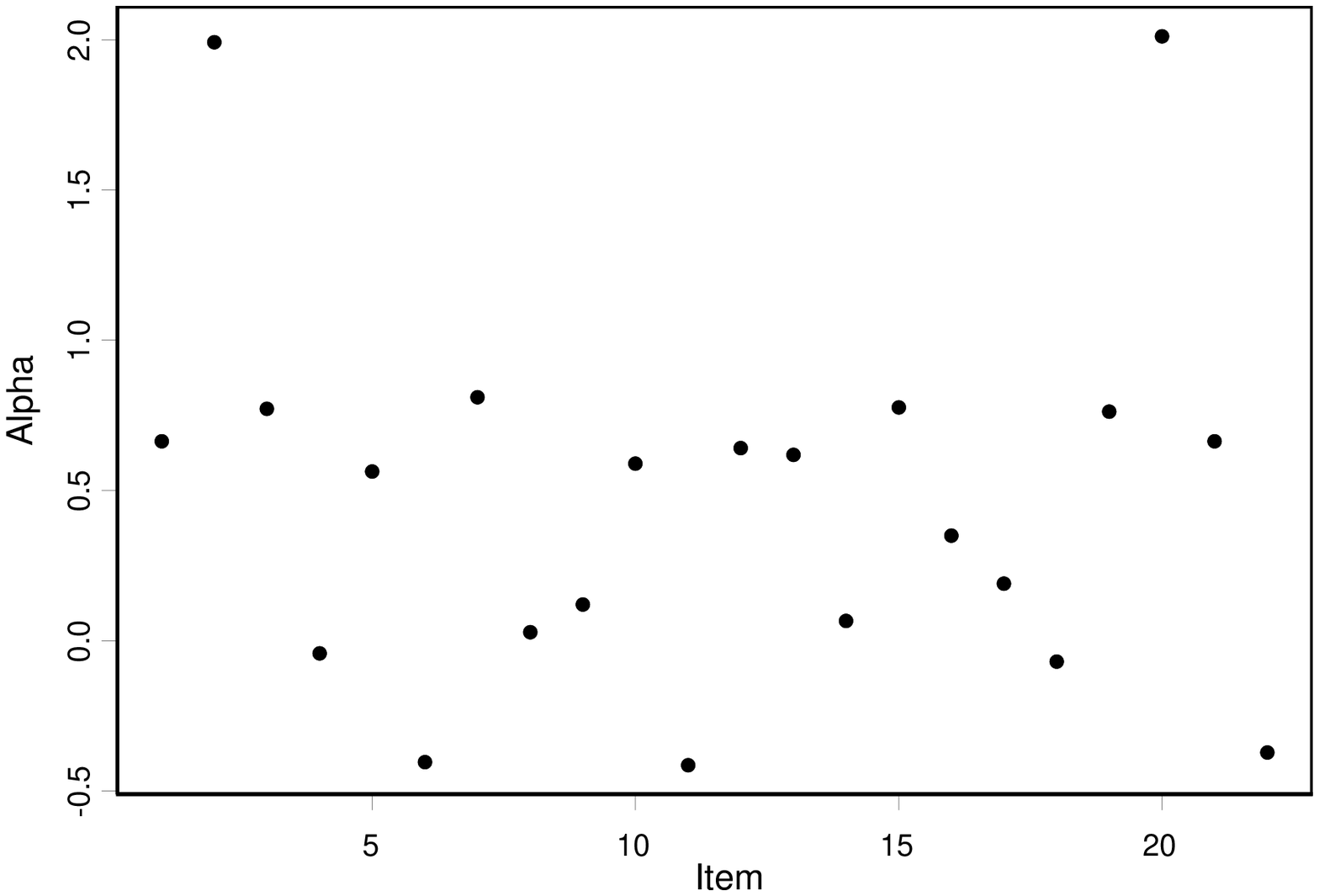}
\includegraphics[scale=0.34]{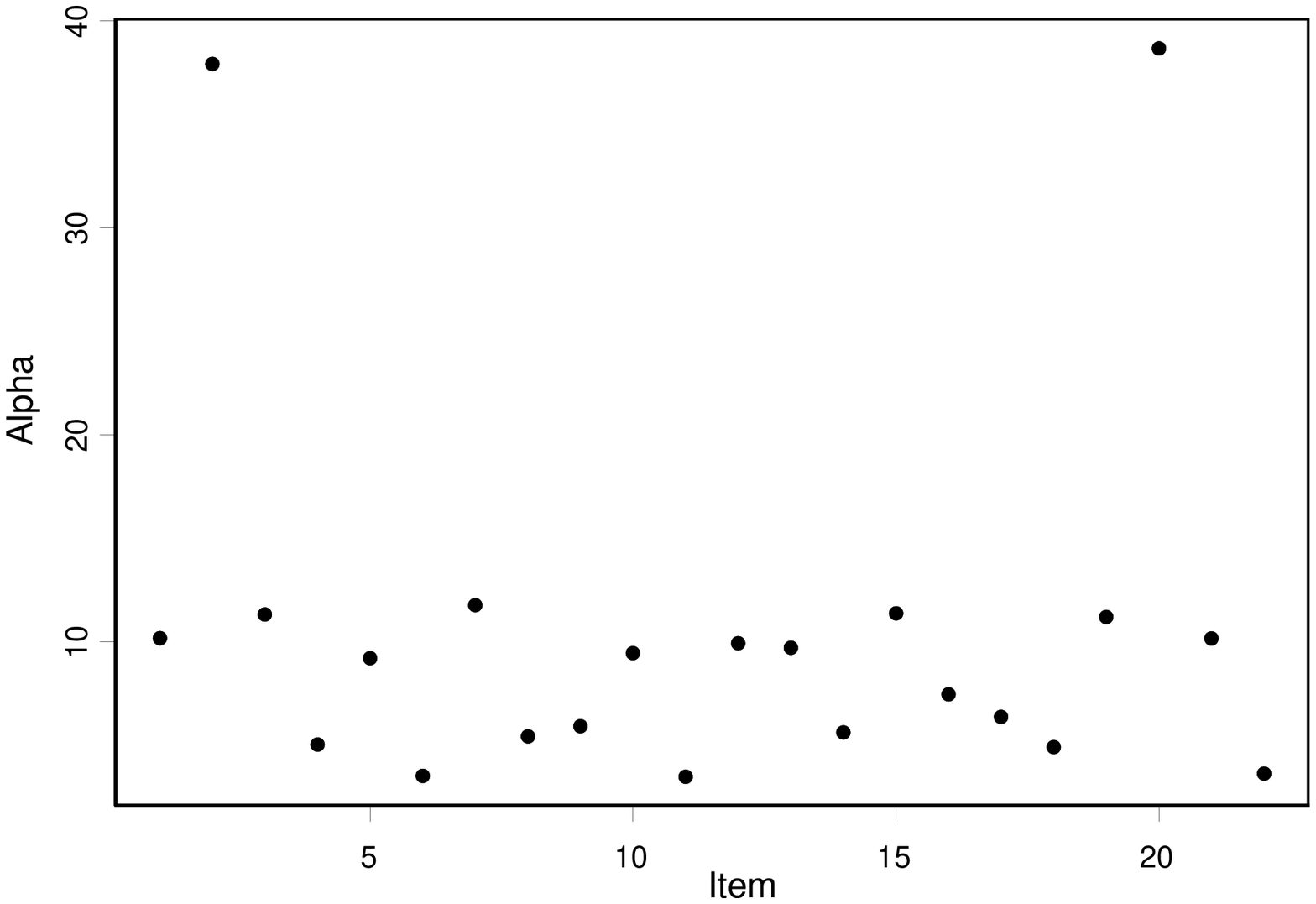}
\includegraphics[scale=0.34]{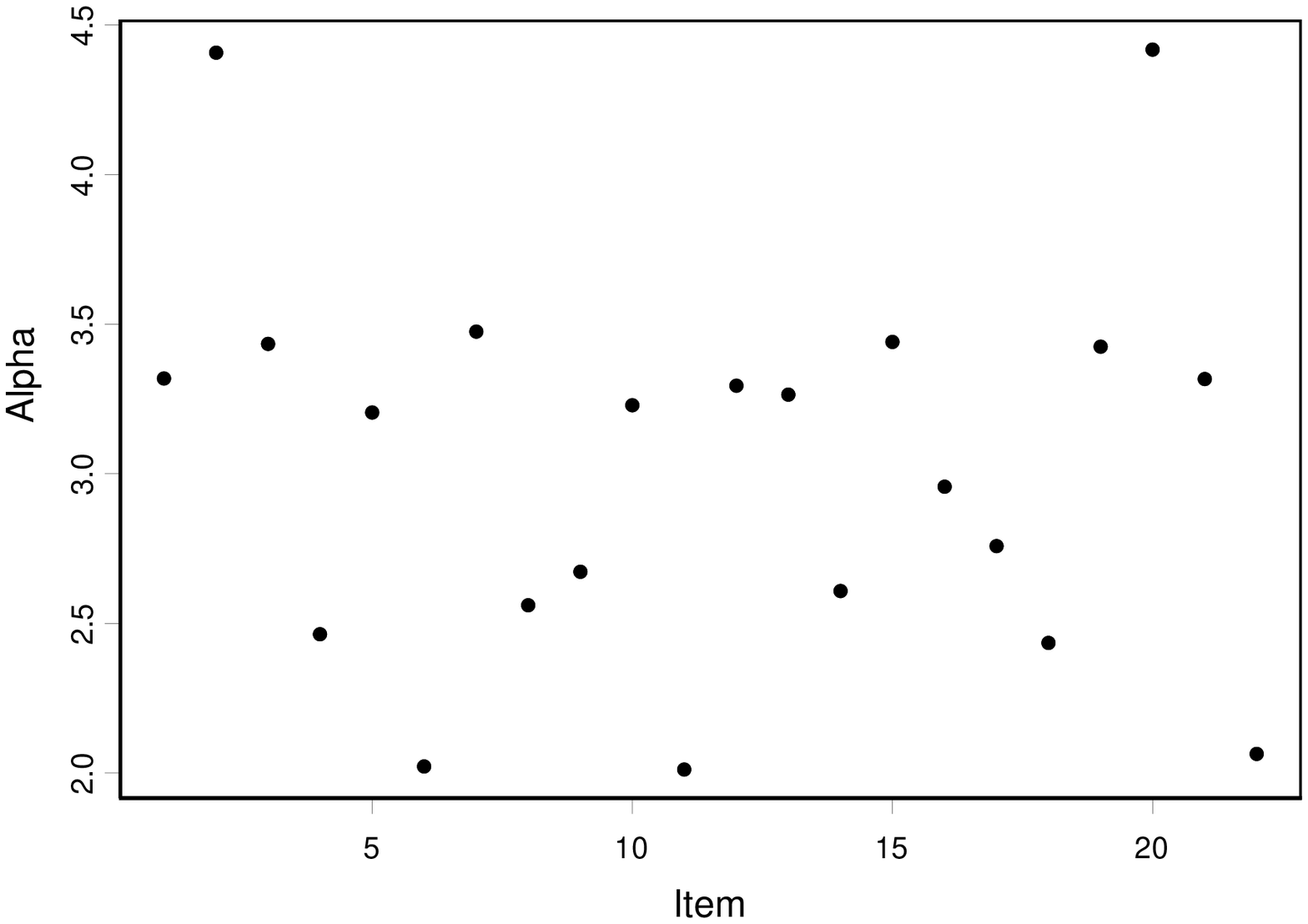}
\caption{Difficulty parameter estimates} \label{g-alpha}
\end{figure}

\begin{figure}[h!]
\centering
\includegraphics[scale=0.3]{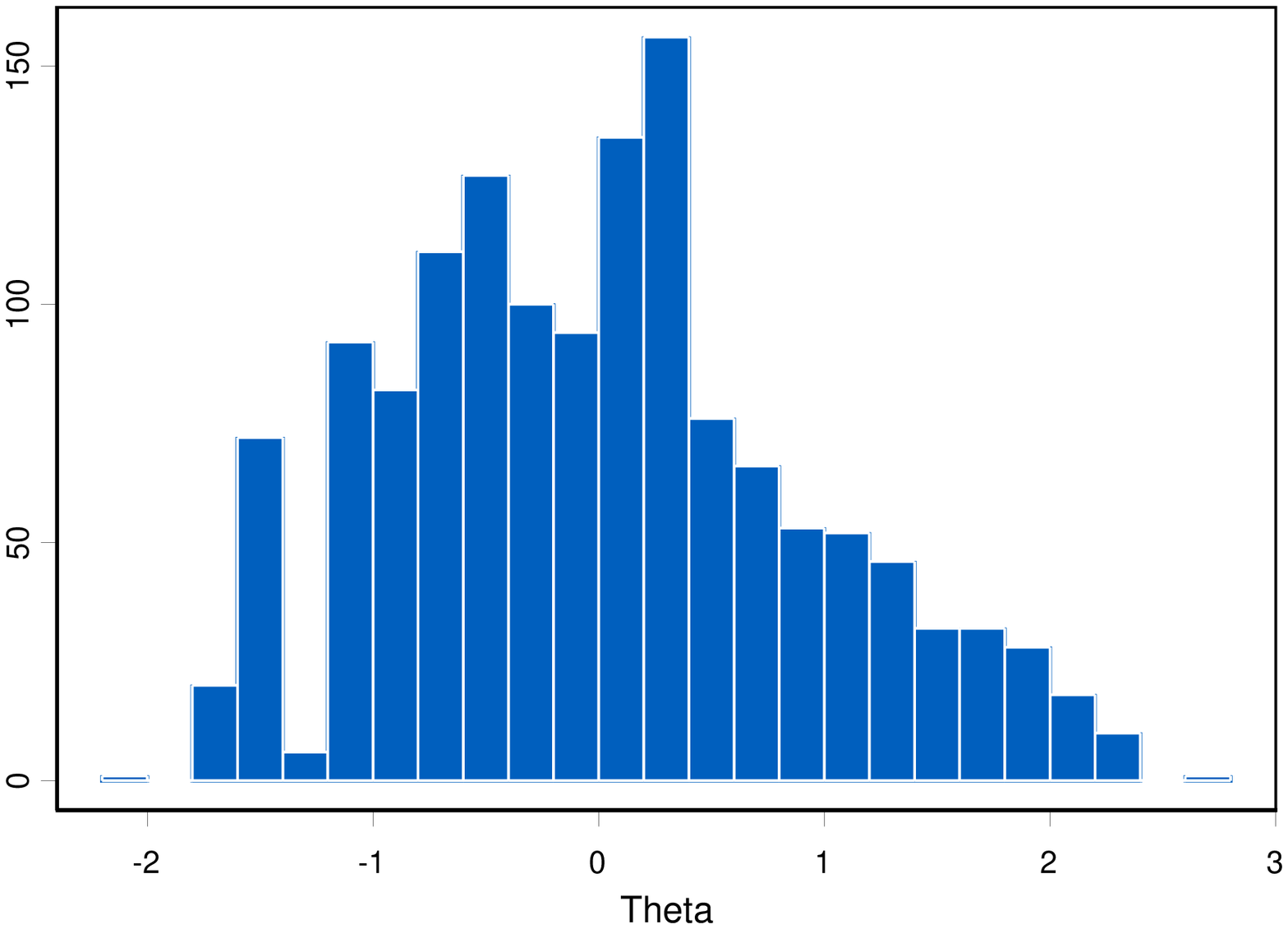}
\includegraphics[scale=0.3]{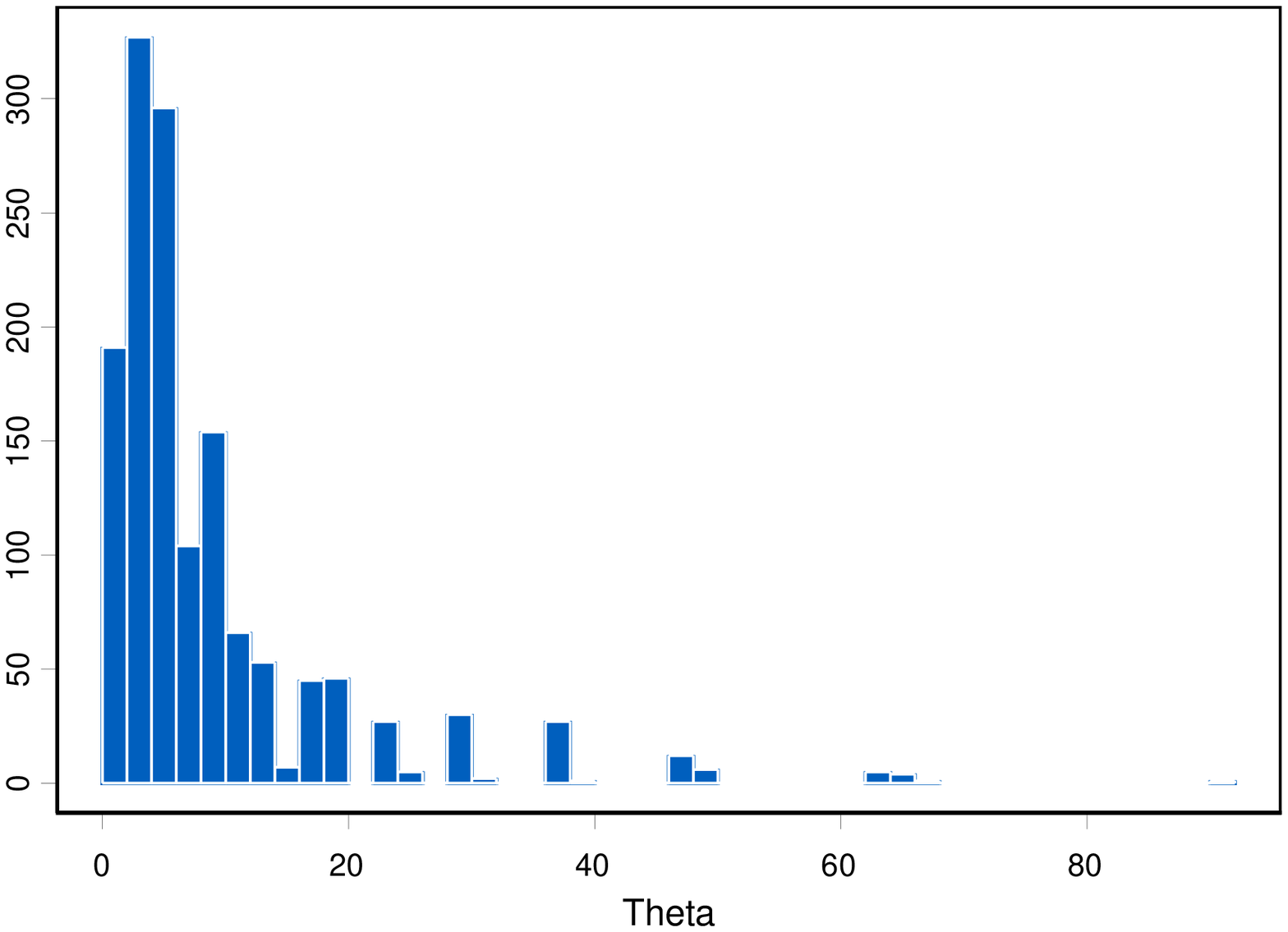}
\includegraphics[scale=0.3]{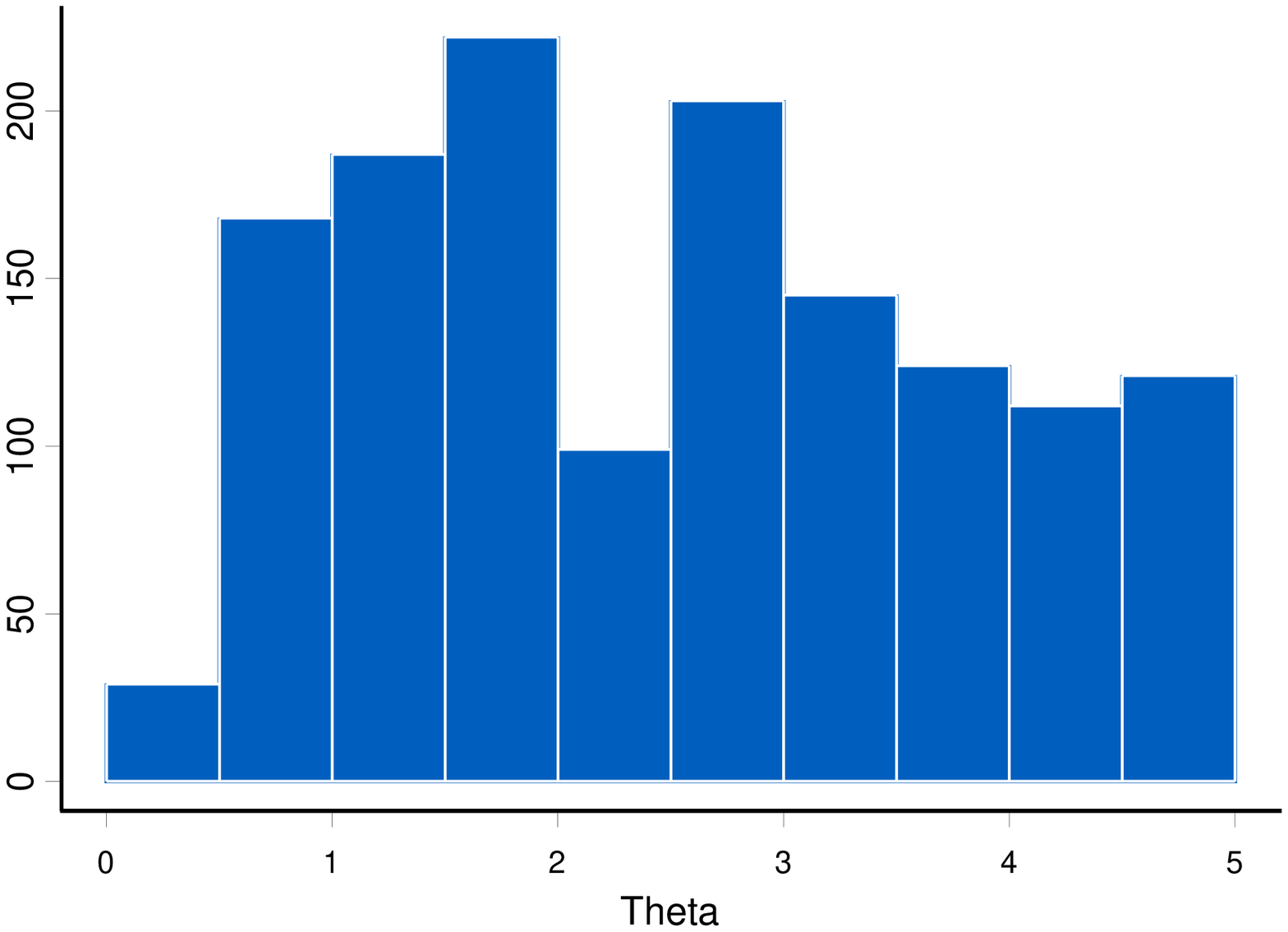}
\caption{Histograms of latent parameter $\theta$}\label{g-histogram}
\end{figure}

Figure \ref{g-alpha} shows the estimates of the difficulty parameter for model \ref{norModel}a (on the left),    where
the level of bullying is assumed to have normal distribution;  for  model \ref{lognorModel}a (on the right),
where the level of bullying is assumed to have lognormal distribution, and for model \ref{UniforModel}a (bottom),
where the level of bullying is assumed to have uniform distribution.
Figure \ref{g-histogram} is the histogram of the latent parameter estimates,  for  model  \ref{norModel}a  on the left,  model \ref{lognorModel}a on the right and model \ref{UniforModel}a at the bottom. For all models, the difficulty parameter estimates follow the same structures.

The parameter estimates were obtained using the free OpenBugs software.
The DIC value for model 1a is $35790.0$, $35740.0$ for model 2a, and $35550.0$ for model 3a. Thus, given that the third model has the smallest DIC value, it is assumed to have the best to fit to the bullying data. It is possible to conclude that the model with bounded range and logit transformation is the best to describe the bullying victimization distribution. The histogram of bulling victimization distributions are given in  Figure \ref{g-histogram}.

\subsection{Anchor of the items and empirical selection levels}

In this section, results of the application of the empirical procedure described in section 3 are presented, to define
bullying victimization levels. From the anchor plots, four groups of items  which define a bullying
victimization scale are defined. These groups are characterized by the following situations:

\begin{enumerate}
\item {\bf Group I}: Items  6, 11 and 22

``They criticize me and blame me for anything that I do or decisions that I make"; ``They make fun of me and play tricks on me", and ``They give me disparaging nicknames." At this level the students did not report that often or sometimes
``They take away my snacks" or `` They force me to do things that endanger my physical integrity".

\item {\bf Group II}: Items: 4, 8, 9, 14, 16, 17 and 18.

``They don't talk to me"; ``They don't count on me for class activities"; ``They maliciously distort all that I say or do";
``They interrupt me continually, impeding  me  expression"; ``They make cruel jokes about my physical appearance";
``They throw objects at me"; `` They hide my possessions"

\item {\bf Group III}: Items: 1, 3, 5, 7, 10, 12 13, 15, 19 and  21.

``They force me to do things that I don't want to do"; ``They persuade others not to talk to me;
``They despise my work, no matter what I do;
``They humiliate me  and ridicule me in front of others "; ``They blame me for everything bad that happens"; ``They threaten me verbally or through gestures"; ``They ignore me and exclude me"; ``They prevent me from communicating";
``They damage my possessions"; and ``They do things to try to get a violent reaction from me".

\item {\bf Group IV}:   Item 2 and  20.

``They force me to do things that endanger my physical integrity" and ``They take away my snacks". The students that are victims of bullying characterized by these two items often are victims of the bullying characterized by each of the other 20 items.

\end{enumerate}

At the first level of harassment, the student is the victim of criticism, ridicule, jokes and derogatory nicknames. In the second, he/she is the victim of isolation in class and phenomena of indirect physical and psychological aggression, also present in level 3, with a greater level of intensity in isolation, becoming a victim of humiliation and threats. At this level, in addition to facing violence, the victim presents a higher level of submission. Finally, in the fourth level, the victim is totally subjected and may be forced to put his/her physical integrity at risk. In conclusion these levels characterize a scale where physical and psychological aggression intensifies with the level of victim submission.

\subsection{IRT regression models}

In this section we present the results of fitting the IRT regression models obtained by assuming that:
\begin{eqnarray}
h(\theta_j)=\beta_0+\beta_1X_{1j} +\beta_2X_{2j}+e_j, \;\; e_i\sim N(0, \sigma^2),
\end{eqnarray}
where $h$ is the identity function in models \ref{norModel}, the logarithm function in model \ref{lognorModel} and the logit function in model
\ref{UniforModel}.
 In this application,  $X_1$ equals 1 for boys and  $0$ for girls, and $X_2$ equals 1 for sixth, seventh and eighth grade, and  $0$ for ninth, tenth and eleventh grade.
In each of analysis it is assumed that the difficulty item parameters $\alpha_i$'s are known,
with the parameter values obtained by fitting the respective IRT model in section \ref{IRTParamEst}.
Thus,  the posterior parameter estimates obtained, assuming  normal  prior distribution $N(0, 0.00001)$ for the regression parameters and gamma prior distribution $G(0.0001, 0.0001)$ for the variance parameter $\sigma^2$, are given by:

\begin{enumerate}
\item {\bf Model 1b}.
Assuming the IRT normal  regression model, the parameter estimates are: $\beta_0=0.052(0.045)$, $\beta_1=-0.043(0.056)$ $\beta_2=-0.138 (0.073)$, $\sigma^2=1.008 (0.050)$. For this model, the DIC value is: $35730.0$. Assuming a model without $X_1$, the parameter estimates and DIC value of the resulting model are given by:  $\beta_0=0.03377 ( 0.03197)$, $\beta_2=-0.1389 (0.06906  )$,
$\sigma^2=1.004(0.049)$. For this model, the DIC value is $35730.0 $

\item {\bf Model 2b}.
Assuming the IRT lognormal regression model, the parameter estimates are: $\beta_0=1.662 (0.112)$, $\beta_1=-0.030(0.069)$, $\beta_2=-1.422 (0.070)$,
$\sigma^2=0.108 (0.005)$. For this model, the DIC value is $35750.0$. Assuming a model without $X_1$, the parameter estimates and DIC value of the resulting model are given by:  $\beta_0= 1.644(0.035)$, $\beta_2=-0.1374(0.074)$,
$\sigma^2=1.006( 0.052)$. For this model, the DIC value is: $ 35750,0$.

\item {\bf Model 3b. \label{EstUnifReg}}
Assuming the IRT uniform regression model, the parameter estimation are: $\beta_0=0.109(0.073)$, $\beta_1= -0.051 (0.117)$, $\beta_2=-0.236(0.118)$,
$\sigma^2=0.327 (0.033)$. For this model, the DIC value is: $35750.0$. Assuming a model without $X_1$, the parameter estimates and DIC value
of the resulting model are given by:  $\beta_0=0.082 (0.059)$, $\beta_2= -0.214 (0.124)$,
$\sigma^2= 0.325 (0.016)$. For this model, the DIC value is: $35750.0 $
\end{enumerate}

From these results of  each of the models,  the same practical conclusion and interpretation can be obtained: there is no bullying victimization difference between girls and boys   and the probability of bullying in grades six, seven and
eight is larger than in grades nine, ten and eleven.

\section{Conclusion}\label{conclusion}

In this paper, new item-response models are proposed assuming that the abilities range from $0$ to $\infty$ or from $0$  to a positive real number $R$. The second interval should be more powerful in the analysis of the school bullying data and seems to be a scale more clearly defined to explain the bullying intensity parameter distribution. This new model presents new theoretical and applied research possibilities. In school bullying studies, it is considered that tests have items with at least three categories of answers: frequently, sometimes and never. Thus, a possible extension of this work is to propose IRT models for graduated response to the analysis of this class of data, assuming that the latent variable ranges in a bounded interval. This is a work in development.

The presented empirical procedure is useful to interpret educational quality tests, in order to understand their theoretical framework, allowing the proposal, for example, of educational policies. The application of the proposed models can be extended to multiples areas of social research, where the framework of the latent variable cannot be assumed to take values from $-\infty$ to $\infty$. In education quality evaluation, for example, the cognitive mathematical achievement or the communicative development can be assumed to be a latent variables ranging in a bounded interval. These models better capture the latent variable distribution and facilitates interpretation and understanding of the results obtained in the analysis.

\section*{Appendix: Bullying Survey}

The bullying victimization test is composed of $22$ items, each of which has three categories related with frequency of the student's victimization described by each of the items. These categories are: {\it Never}, {\it Sometimes} and {\it Frequently}. The lowest category {\it Never} is related to absence of bullying.

\begin{enumerate}
\item They force me to do things that  I don't want to do.
\item They force me to do things that endanger my physical integrity.
\item They persuade others not to talk to me.
\item They don't talk to me.
\item  They despise my work, no matter what I do.
\item  They criticize me and blame me for anything that I do or decisions that I take.
\item  They humiliate me  and ridicule me in front of others.
\item They don't count on me for class activities.
\item They maliciously distort all that I say or do.
\item They blame me for everything bad that happens.
\item They make fun of me and play tricks on me.
\item They threaten me verbally or through gestures.
\item They ignore me and exclude me.
\item They interrupt me continually, impeding  me  expression.
\item They prevent me from communicating.
\item They make cruel jokes about my physical appearance.
\item  They throw objects at me.
\item They hide my possessions.
\item They damage my possessions.
\item They take away my snacks.
\item They do things to try to get  a violent reaction from me.
\item They give disparaging nicknames.
\end{enumerate}

\section{References}

Andrade, D., F. Tavares H., and Cunha R.(2000). Teoria da resposta ao item: conceitos e aplicac\~{o}es. S\~{a}o Pablo-Brasil ABE- Associa\c{c}\~{a}o Brasileira de Estat\'istica.

Bartholomew, D., J. (1980). Factor analysis for categorical data. Journal of the Royal Statistical Society, 42(3), 293-321.

Bartholomew, D., J. (1984). The foundations of factor analysis. Biometrika, 71(2), 221-232.

Beaton, A., E., and  Allen, N., L. (1992).  Interpreting scales through scale anchoring. Journal of Educational and Behavioral Statistics, 17(2), 191-204.

Beaton, A., E., and Johnson, E., G. (1992). Overview of the scaling methodology used in the national assessment.
Journal of Educational Measurement, 29(2), 163-175.

Birnbaun, A. (1968).  Some latent trait models and their use in inferring an examinee's ability. In F. M. Lord \& M.R. Novick (Eds.).  Statistical theories of mental tests score (pp. 395-479). Reading, MA. Addison-Wesley.

Bock, R., D., and Aitkin, M. (1981). Marginal maximum likelihood estimation of item parameters: An applications of a EM algorithm.  Psychometrika,  46, 179-197.

Bock, R., D., and Lieberman, M. (1970).  Fitting a response model for n dichotomously scored item. Psychometrika, 35, 179-197.

Cepeda-Cuervo, E., and Caicedo-S\'anches, G. (2012). Acoso escolar: caracterizaci\'on, consecuencias y prevenci\'on. OEA: Organizacion de los Estados Americanos, 1-7.

Cervantes, V. H. Cepeda, E., and Camargo, S. L. (2008). Una propuesta para la obtención de niveles de desempeño en los modelos de teor\'ia de respuesta al \'item. {\it Avances en Medici\'on}, 6(1), 49-58.

Cepeda-Cuervo, E. Pacheco-Dur\'an, P. N. Garc\'ia-Barco, L., and Piraquive-Pe\~{n}a, C. J. (2008). Acoso escolar a estudiantes de educaci\'on
b\'asica y media.  Revista de salud p\'ublica, 10 (4), 517-528.

Fisher, G., H. (1995). Some neglected problems in IRT.  Psychometrika, 60, 449-487.

Goodboy, A., K., and Martin, M., M. (2015). The personality profile of a cyberbully: Examining the Dark Triad. Computers in human behavior, 49, 1-4.

Hambleton, R. K. Swaminathan, H., and  Rogers, H. J. (1991). Fundamentals of item response theory.  California: Sage.

Harris, D. (1989). Comparison of 1-, 2-, and 3-Parameter IRT Models. Educational Measurement: Issues and Practice, 8(1), 35-41.

Lazarsfeld, P., F. (1955). Recent developments in latent structure analysis. Sociometry, 18(4), 391–403.

Lord, F., M. (1950). A theory of test scores. Psychometric Monographs, 7. Chicago: University of Chicago Press

Lord, F., M. (1980). Applications of item response theory to practical testing problems. Hillsdale, New Jersey:
Lawrence Erlbaum Associates.

Ludlow, L., H. and Haley, S., M. (1995). Rasch model logits: Interpretation, use, and transformation. Educational and Psychological Measurement, 55(6), 967-975.

Olweus, D. (1994). Bullying at school. In Aggressive behavior (pp. 97-130). Springer, Boston, MA.

O\~{n}ate, A., and Pi\~{n}uel, I. (2005). Informe Cisneros VII: Violencia y acoso escolar en alumnos de primaria, ESO y bachiller. Madrid:  Instituto de Innovación educativa y Desarrollo directivo.

Mu\~{n}iz, J. (1997). Introducción a la teoría de respuesta a los ítems. Madrid: Ediciones Pirámide.

Rasch, G. (1993). Probabilistic models for some intelligence and attainment tests. MESA Press, 5835 S. Kimbark Ave., Chicago, IL 60637; e-mail: MESA@ uchicago. edu; web address: www. rasch.org; tele..

Spiegelhalter, D. Thomas, A. Best, N., and  Lunn, D. (2003). WinBUGS user manual.

Van der Linden, W., J., and Hambleton, R. K. (1997). Item response theory: Brief history, common models, and
extensions. En W. J. van der Linden \& R. K. Hambleton (Eds.) Handbook of modern item response theory (pp.
1-28). New York: Springer-Verlag.

Wilson, M. (1989). Saltus: A psychometric model of discontinuity in cognitive development. Psychological Bulletin,105(2), 276.

\end{document}